\documentclass[
aps,prl,twocolumn,reprint,superscriptaddress,nofootinbib,floatfix
]{revtex4-2}

\usepackage{amsmath,amssymb,bm}
\usepackage{graphicx}
\usepackage{hyperref}
\hypersetup{hidelinks}
\usepackage{color}

\begin{document}

\title{Evidence for Quartet Binding of Valence Neutrons in $^8$He}

\author{Young-Ho Song}
\affiliation{
Institute for Rare Isotope Sciences,
Institute for Basic Science,
Daejeon 34000, Korea
}

\author{Yuan-Zhuo Ma}
\affiliation{
Facility for Rare Isotope Beams and
Department of Physics and Astronomy,
Michigan State University,
East Lansing, MI 48824, USA
}

\author{Dean Lee}
\affiliation{
Facility for Rare Isotope Beams and
Department of Physics and Astronomy,
Michigan State University,
East Lansing, MI 48824, USA
}

\date{\today}

\begin{abstract}
Multimodal neutron superfluidity predicts quartets formed as bound states of
two spin-singlet $s$-wave neutron pairs. We present evidence that the four
valence neutrons of $^8$He realize the finite-system analogue. Quartet binding
depends not only on the strength of neutron-neutron attraction but also on the
number and symmetry of sufficiently strong attractive pair modes. Pauli
blocking limits reuse of the same pair structure, while additional modes can
provide extra binding. A partial-wave analysis of the neutron-neutron interaction in Daejeon16
no-core shell-model calculations identifies cooperative $^1S_0$ pairing
and additional $^3P_2$ attraction as the dominant neutron-neutron
contributions to this nonadditive binding, while density cumulants reveal
connected four-neutron correlations.
\end{abstract}

\maketitle

Among light nuclei, $^8$He is singular. It is the only neutron-rich nucleus where the binding of neutron pairs increases with neutron number. The two-neutron separation
energy rises from $S_{2n}(^6\mathrm{He})=0.975$~MeV to
$S_{2n}(^8\mathrm{He})=2.125$~MeV. Whatever holds the four valence neutrons
together is therefore not one pair binding mechanism applied twice. It is tempting to
ascribe the extra binding of the second pair simply to the $0p_{3/2}$ subshell
closure at $N=6$ neutrons. As we shall see, restricting the valence neutrons
to $0p_{3/2}$ already produces the negative three-point curvature associated
with the extra binding of the second pair, but it does not provide enough
absolute binding to bind the valence neutrons in either $^6$He or $^8$He.
Additional correlations are therefore essential. In the full \emph{ab initio}
wave function, irreducible multi-subshell interaction contributions and
connected four-neutron cumulants show that higher-subshell configurations
participate importantly. What microscopic interactions organize
this correlated structure? An answer is suggested by multimodal superfluidity,
recently identified in high-fidelity neutron matter calculations in uniform
matter with attractive $s$- and $p$-wave interactions \cite{Ma:2026dti}. Alongside ordinary $s$-wave pairs and
entangled double-$p$-wave pairs, this phase contains quartets that can be
viewed as bound states of two $s$-wave pairs. 

The four valence neutrons of
$^8$He offer a particularly simple finite system in which to examine the
mechanism producing extra binding for the second valence pair.  To organize the resulting pair correlations, we use the standard
pair-multipole decomposition of the two-body interaction
\cite{Lane:1964NuclearTheory,RingSchuck:1980NuclearManyBody}.  We also introduce a concept called attractive rank, which counts the number of attractive pairing modes in the interaction kernel that exceed some threshold strength.  The calculations of Ref.~\cite{Ma:2026dti} rest on
nuclear lattice effective field theory (NLEFT)
\cite{Lee:2004NLEFT,Borasoy:2007NLEFT,Epelbaum:2010NLEFT,Lahde:2014NLEFT,Elhatisari:2016NLEFT,Lu:2019NLEFT,Elhatisari:2024Wavefunction,Lee:2025NLEFTReview,Song:2025ofd}. Here we test the
quartet mechanism with an independent \emph{ab initio} method that works in the continuum rather than the lattice.

Neutron pairing, halo structure, and continuum correlations in neutron-rich
helium have been studied with quantum Monte Carlo, no-core shell-model,
cluster, complex-energy configuration-interaction, and open-quantum-system
approaches
\cite{Wiringa:2000A8,Caurier:2006Helium,KanadaEnyo:2007He8,Hagen2007Complex,Papadimitriou:2011Helium,Michel:2003GSM,Fossez:2018Helium,Wang:2021PairDynamics}.
Gamow-shell-model and complex-energy calculations including nonresonant
continuum states find strong configuration mixing and a strongly mixed $^8$He
ground state rather than an inert $(0p_{3/2})^4$ configuration
\cite{Michel:2003GSM,Fossez:2018Helium}, and $^{4}$He+4n core-plus-valence
\cite{Hagino:2008vm} and microscopic cluster \cite{Yamaguchi:2023xsx}
calculations likewise find substantial departures from that configuration
with appreciable two-dineutron components. Recent NLEFT studies of
multineutron correlations in light nuclei also find predominantly
dineutron--dineutron configurations in $^8$He, with only a small compact
four-neutron component \cite{Zhang:2025Multineutron}. Our aim is not to attribute
the pattern of nuclear binding uniquely to pair--pair attraction nor to replace other
continuum studies.  We purpose instead is to use the same \emph{ab initio} Hamiltonian for
$^4$He, $^6$He, and $^8$He and resolve which microscopic components of the
neutron-neutron interaction generate the nonadditive binding pattern.

The multimodal \emph{neutron quartet} is distinct from the
\emph{tetraneutron}: the latter is a four-neutron state in vacuum, the former
a connected four-neutron correlation in a many-body environment
\cite{Ma:2026dti}. The near-threshold structure observed in
$^8$He$(p,p\alpha)4n$ \cite{Duer:2022Tetraneutron} admits several
interpretations
\cite{Hiyama:2016Tetraneutron,Fossez:2017Tetraneutron,Higgins:2020Tetraneutron,Lazauskas:2023FourNeutron};
Ref.~\cite{Lazauskas:2023FourNeutron} reproduces it with an $S$-wave-only
neutron-neutron interaction, a final-state mechanism distinct from the
quartet mechanism considered here. Lattice calculations of the free
four-neutron system likewise find no resonance plateau
\cite{Wu:2026TetraneutronLattice}.

For an operator $O$, we define the three-point difference
\begin{equation}
X_O=
O(^8\mathrm{He})
-2O(^6\mathrm{He})
+O(^4\mathrm{He}).
\label{eq:XH}
\end{equation}
This quantity is the standard two-neutron gap (second mass difference) for Hamiltonian
\begin{equation}
X_H=S_{2n}(^6\mathrm{He})-S_{2n}(^8\mathrm{He})\equiv\delta_{2n},
\end{equation}
centered on $^6$He \cite{Dobaczewski:1995ClosedShells,Bender:2002Z82,Scalesi:2024Correlations}.
Such mass filters are sensitive in general to shell structure, deformation,
continuum coupling, and other many-body correlations, so $X_H$ alone is not a
unique signature of quartet binding. For two independent valence pairs coupled
to an unchanged environment, however, we find that $X_H=0$. A nonzero value therefore
quantifies nonlinearity with respect to neutron number along the isotope chain. Our purpose is to resolve the
microscopic origin of that nonlinearity. Experiment gives
$X_H^{\rm exp}=-1.1496$ MeV.

\begin{figure}[t]
\centering
\includegraphics[width=\columnwidth]{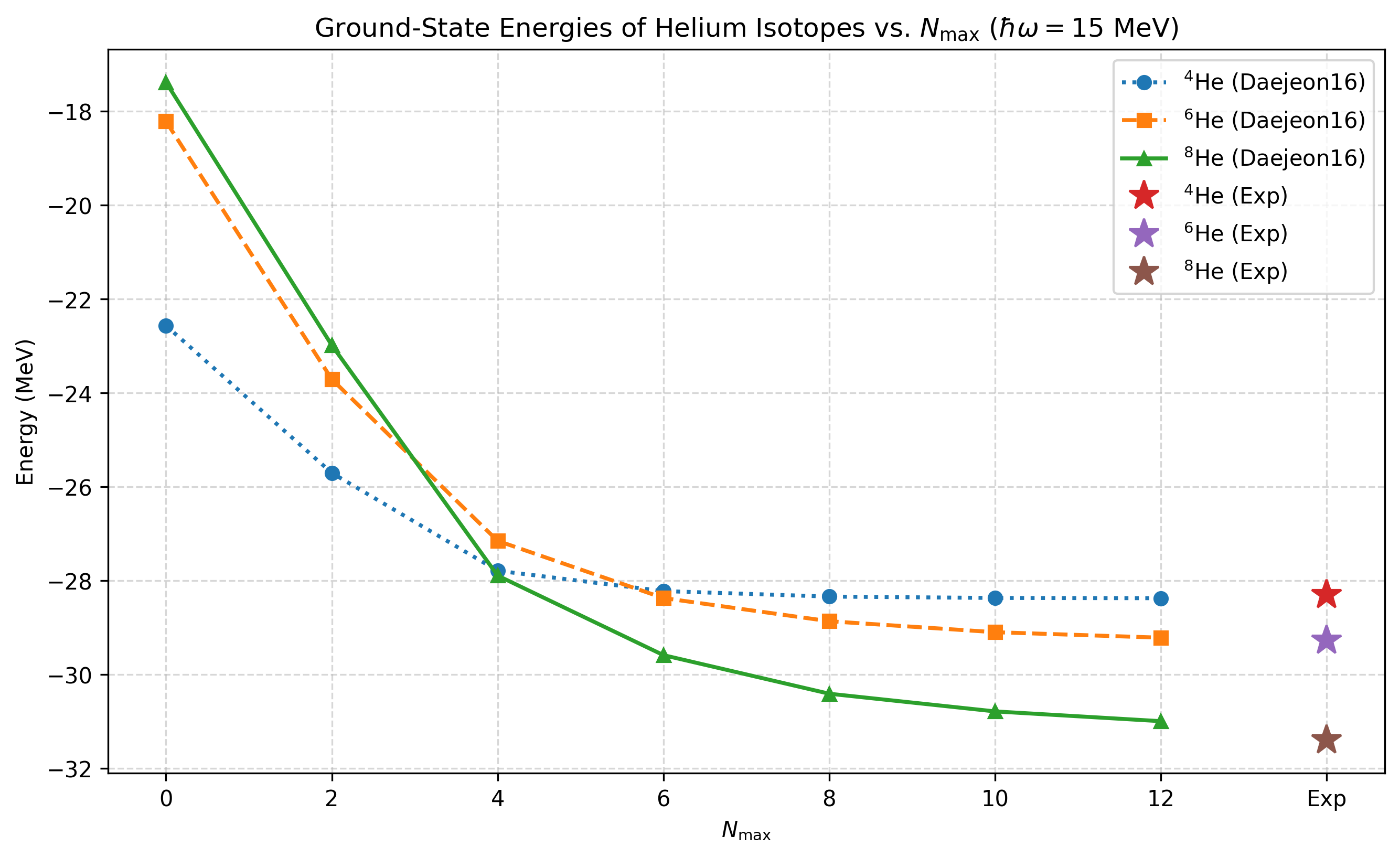}
\caption{
Daejeon16 ground-state energies at $\hbar\omega=15$ MeV.
The calculated $^8$He energy falls below the separately calculated
$^4$He energy at $N_{\max}=4$, whereas the $^6$He energy does so only at
$N_{\max}=6$. Experimental energies are shown at right.
}
\label{fig:he_convergence}
\end{figure}

\emph{Ab initio evidence.---}
We calculate $^4$He, $^6$He, and $^8$He in the no-core shell model (NCSM)
using Daejeon16 \cite{Shirokov:2016Daejeon16,Kim:2019gnl} with Coulomb interaction. 
The nuclear interaction
matrix elements were generated using NuHamil \cite{Miyagi2023NuHamil}, and
the no-core shell-model calculations were performed with BIGSTICK
\cite{Johnson2013Factorization,Johnson2025BIGSTICK}.
Daejeon16 starts from an SRG (similarity renormalization group) evolved chiral N3LO NN interaction and applies phase-equivalent transformations adjusted to light nuclei. It provides a good description of many observables in light nuclei without explicit three-nucleon forces.
Convergence in a
harmonic-oscillator basis is demanding because the halo structure and nearby
continuum require large model spaces
\cite{Caurier:2006Helium,Papadimitriou:2011Helium,Fossez:2018Helium,Lazauskas:2023FourNeutron}.
We use Daejeon16 because its comparatively rapid convergence makes
high-$N_{\max}$ calculations of $^6$He and $^8$He computationally feasible.
The harmonic-oscillator basis does not treat scattering asymptotics explicitly,
so these calculations are not a substitute for a continuum formulation.
Our conclusions concern the Daejeon16 NCSM calculation and the decomposition of
its energy differences, and should not be interpreted as demonstrating
interaction independence. Figure~\ref{fig:he_convergence} shows the ground-state
energies versus $N_{\max}$ at $\hbar\omega=15$ MeV.

The convergence displays a notable hierarchy. At $N_{\max}=4$,
\begin{equation}
E(^8\mathrm{He})-E(^4\mathrm{He})
=
-0.103~\mathrm{MeV},
\end{equation}
so the calculated $^8$He energy already lies below the separately calculated
$^4$He energy. At the same truncation,
\begin{equation}
E(^6\mathrm{He})-E(^4\mathrm{He})
=
+0.637~\mathrm{MeV},
\end{equation}
and the $^6$He energy remains above it, crossing below only at
$N_{\max}=6$. Thus within the same truncated variational space the
correlations gained as the model space is enlarged are especially favorable
for the four-valence-neutron system. The same hierarchy occurs at
$\hbar\omega=25$ MeV, where $^8$He lies below $^4$He already at
$N_{\max}=10$ while $^6$He does so only at $N_{\max}=12$. Numerical
results for both oscillator frequencies are given in the Supplemental
Material.

The three-point difference itself is not informative in the smallest
spaces. At $N_{\max}=0$ the four valence neutrons are unbound relative to the
core, yet $X_H$ is large and negative, because $^6$He is described even more
poorly there than $^8$He; the mass filter registers the relative quality of
the truncation rather than a converged mechanism. Only for $N_{\max}\ge6$,
once both isotopes are bound relative to $^4$He, does $X_H$ settle. At larger
$N_{\max}$, the three-point difference in Eq.~(\ref{eq:XH}) converges toward
approximately $-1$ MeV. At $\hbar\omega=15$ MeV we obtain
$X_H=-1.019$ MeV at $N_{\max}=8$ and $-0.962$ MeV at $N_{\max}=10$,
compared with the experimental value
$X_H^{\rm exp}=-1.1496$ MeV. Thus the enhanced binding associated with the
second valence-neutron pair is already stable in the largest directly
calculated spaces.

\emph{Large kinetic--interaction cancellation.---}
The small final value of $X_H$ should not be confused with a weak underlying
interaction effect. Writing the intrinsic Hamiltonian as
$H=T_{\rm rel}+V$, the $N_{\max}=10$ wave functions give
\begin{equation}
X_{T_{\rm rel}}=+10.817~\mathrm{MeV},
\qquad
X_V=-11.780~\mathrm{MeV}.
\label{eq:energy_balance}
\end{equation}
Using the unrounded values, these contributions sum to
$X_H=-0.962$ MeV. The observed nonadditive binding is therefore the small
residual of a negative interaction contribution more than an order of magnitude
larger than the final $X_H$ and a large positive intrinsic-kinetic-energy
contribution.

After subtraction of the center-of-mass kinetic energy,
\begin{equation}
T_{\rm rel}
=
\sum_i\frac{\bm p_i^2}{2m}
-\frac{\bm P_{\rm cm}^2}{2Am}
=
\frac{1}{2Am}
\sum_{i<j}
(\bm p_i-\bm p_j)^2.
\label{eq:Trel_pair}
\end{equation}
Although Eq.~(\ref{eq:Trel_pair}) permits a formal decomposition into
neutron-neutron, proton-neutron, and proton-proton pair terms, the factor
$1/A$ and the numbers of each type of pair change along the isotope chain.
We therefore keep $X_{T_{\rm rel}}$ as a full-system quantity and resolve
only the potential contribution by pair type.

The potential contribution to the three-point difference separates as
\begin{equation}
\begin{aligned}
X_{nn,V}&=-5.795~\mathrm{MeV},\\
X_{pn,V}&=-5.187~\mathrm{MeV},\\
X_{pp,V}&=-0.797~\mathrm{MeV}.
\end{aligned}
\label{eq:V_pair_balance}
\end{equation}
The neutron-neutron and proton-neutron terms both make substantial negative
contributions. The reason for focusing on the neutron-neutron interaction is
particularly transparent in the restricted $0\hbar\Omega$ configurations
$(0s_{1/2})^4(0p_{3/2})^n$, with $n=0,2,4$ for $^{4,6,8}$He. With the core
fixed, core-core contributions are constant, while core-valence interactions
and the intrinsic harmonic-oscillator kinetic energy are affine functions with respect to $n$.
Their three-point differences therefore vanish exactly.
For the valence-valence neutron interaction, the first pair contributes the
$J=0$ matrix element $V_0$, while the filled $(0p_{3/2})^4$ configuration has
interaction energy $V_0+5V_2$, where $V_2$ is the corresponding $J=2$
matrix element. Since
$X_{T_{\rm rel}}^{(0p_{3/2})}=X_{pn,V}^{(0p_{3/2})}
=X_{pp,V}^{(0p_{3/2})}=0$, we obtain
\begin{equation}
X_H^{(0p_{3/2})}
=
X_{nn,V}^{(0p_{3/2})}
=
5V_2-V_0
=
-4.634~\mathrm{MeV}.
\label{eq:p32_restricted_balance}
\end{equation}
This is the same $V_0$--$V_2$ construction used later in our analysis to illustrate the definition of attractive rank and worked out in detail in the Supplemental
Material, Sec.~\ref{sec:p32_analysis}. Thus the valence-neutron interaction
already generates the negative three-point curvature in this simplest
restricted description. This should not be confused with sufficient absolute
binding: with the valence neutrons confined to $0p_{3/2}$, both $^6$He and
$^8$He remain unbound. The nonzero proton-neutron contribution in the full
NCSM calculation therefore reflects nonlinear core response and configuration
mixing absent in this restricted space calculation.

\begin{figure}[t]
\centering
\includegraphics[width=\columnwidth]
{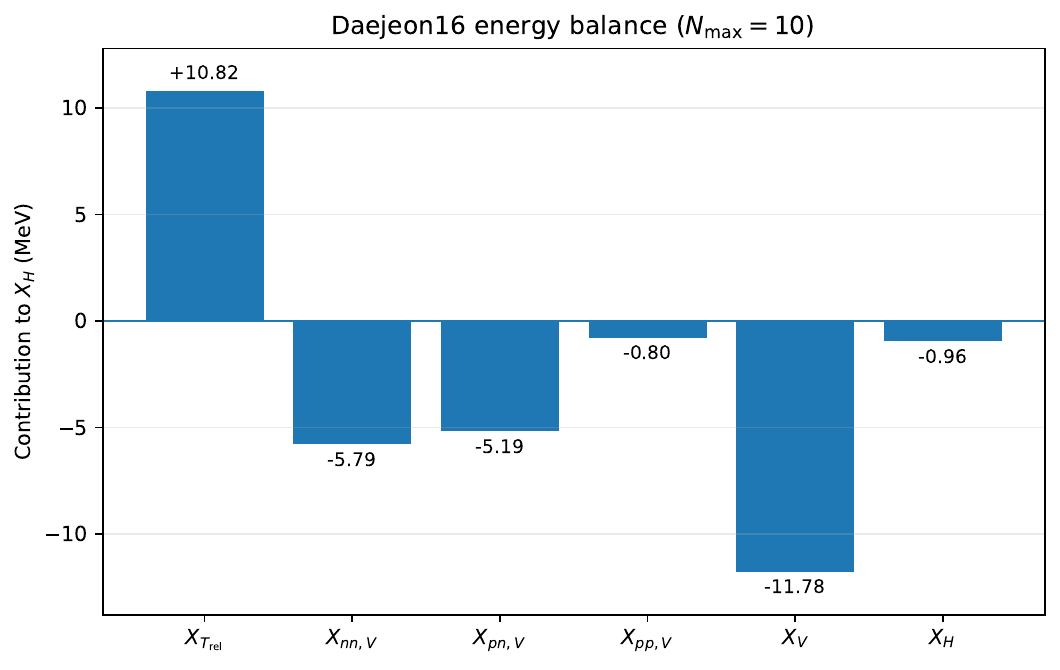}
\caption{
Daejeon16 energy balance at $N_{\max}=10$.
$X_{T_{\rm rel}}$ and $X_H$ are shown only as totals;
$X_V$ is also resolved into neutron-neutron, proton-neutron,
and proton-proton terms.
}
\label{fig:energy_balance}
\end{figure}

The energy balance establishes strong interaction-driven nonlinearity with respect to neutron number but
does not identify which neutron-neutron channels generate it. We therefore
resolve the pure neutron-neutron interaction into subshell clusters and
partial waves. For a set of subshells $S$, let $P_S$ project onto
configurations constructed from $S$ and define
\begin{align}
E_{\mathcal O}(S)\label{eq:expect}
&=
\langle\Psi|P_S\mathcal O P_S|\Psi\rangle,
\\
\Delta_{\mathcal O}(S)\label{eq:in_ex}
&=
\sum_{T\subseteq S}
(-1)^{|S|-|T|}
E_{\mathcal O}(T).
\end{align}
The alternating sum implements inclusion--exclusion: contributions already
present in any proper subset of $S$ cancel, leaving the irreducible
contribution that requires all subshells in $S$.
For example, if $T_1,T_2,T_3$ denote three
distinct subshells, then Eq.~\eqref{eq:in_ex} gives
\begin{align}
&\Delta_{\mathcal O}(T_1\cup T_2\cup T_3) ={}E_{\mathcal O}(T_1\cup T_2\cup T_3) \nonumber\\
&-E_{\mathcal O}(T_1\cup T_2)
-E_{\mathcal O}(T_1\cup T_3)
-E_{\mathcal O}(T_2\cup T_3) \nonumber\\
&+E_{\mathcal O}(T_1)+E_{\mathcal O}(T_2)+E_{\mathcal O}(T_3)
-E_{\mathcal O}(\emptyset).
\end{align}
We now also define
\begin{equation}
X_{\mathcal O}(S)
=
\Delta_{\mathcal O}(S;{}^8\mathrm{He})
-2\Delta_{\mathcal O}(S;{}^6\mathrm{He})
+\Delta_{\mathcal O}(S;{}^4\mathrm{He}).
\label{eq:cluster_difference}
\end{equation}
This second subtraction removes contributions additive in the number of
valence pairs.

\begin{figure}[t]
\centering
\includegraphics[width=\columnwidth]
{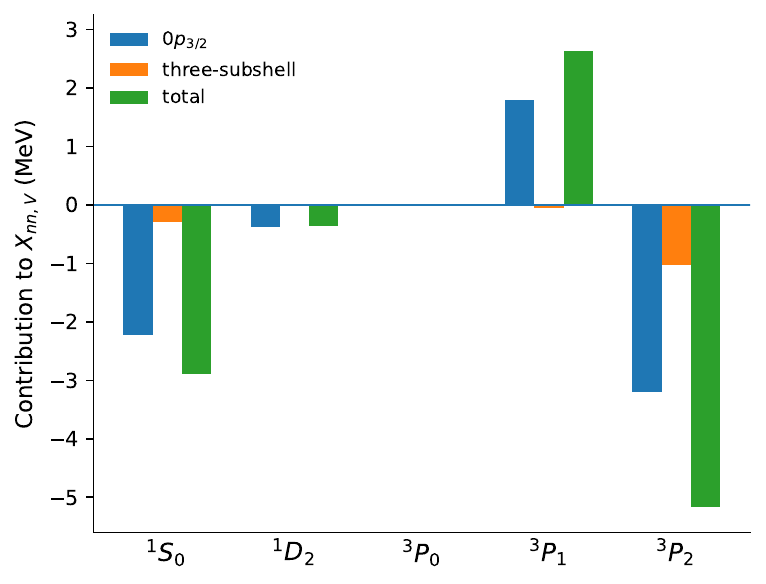}
\caption{
Partial-wave decomposition of $X_{nn,V}$ at $N_{\max}=10$.
Shown are the isolated $0p_{3/2}$ contribution, the irreducible contribution
summed over all distinct three-subshell sets, and the total contribution
summed over all cluster sizes. The $^1S_0$ and $^3P_2$ channels provide
substantial attraction overall, while $^3P_1$ provides substantial
repulsion. The irreducible three-subshell attraction is dominated by
$^3P_2$, with an additional attractive contribution from $^1S_0$.
}
\label{fig:partial_wave}
\end{figure}

Figure~\ref{fig:partial_wave} shows the $N_{\max}=10$ decomposition.
For the isolated $0p_{3/2}$ contribution, the partial-wave contributions
$(^1S_0,^1D_2,^3P_0,^3P_1,^3P_2)$ are
\begin{equation}
(-2.217,-0.370,\,0.000,+1.791,-3.204)~\mathrm{MeV}.
\label{eq:p32_partial_waves}
\end{equation}
For comparison, the same restricted $0p_{3/2}$ model used in
Eq.~(\ref{eq:p32_restricted_balance}) gives a complementary shell-model view
in terms of the $J=0$ and $J=2$ matrix elements $V_0$ and $V_2$. Their
partial-wave decomposition separates the interaction contributions from
adding the first and second valence-neutron pairs; details are given in the
Supplemental Material, Sec.~\ref{sec:p32_analysis}.
The sum of irreducible contributions over all three-subshell sets gives the values
\begin{equation}
(-0.295,+0.015,-0.004,-0.047,-1.034)~\mathrm{MeV}.
\label{eq:three_subshell_partial_waves}
\end{equation}
The sum over all possible cluster sizes gives the total partial-wave
contributions
\begin{equation}
(-2.884,-0.367,+0.022,+2.632,-5.171)~\mathrm{MeV}.
\label{eq:total_partial_waves}
\end{equation}
These five channels account for $-5.769$ MeV of the full
$X_{nn,V}=-5.795$ MeV; all remaining partial-wave contributions together
amount to only $-0.026$ MeV. Thus $^1S_0$ and $^3P_2$ both provide substantial
attraction, while $^3P_1$ provides substantial repulsion. The $^3P_2$
interaction is the largest attractive contribution both in the total
decomposition and in the irreducible three-subshell sector. Since the latter
is evaluated with the projected full NCSM wave functions, it shows that an
important part of the $^3P_2$ attraction requires configurations spanning
multiple subshells rather than only a pure $(0p_{3/2})^4$ configuration.

The partial-wave labels used here, such as $^1S_0$, $^1D_2$, and
$^3P_J$, denote neutron-neutron relative-motion partial waves specified by
$L$, $S$, and $J$, not occupations of particular single-particle shells.
In the NCSM they are represented by harmonic-oscillator configurations
spanning multiple shells; scattering asymptotics are not treated explicitly.
Thus, the $^3P_2$ contribution denotes the relative-motion $L=1$, $S=1$,
$J=2$ partial wave, not a pair confined to the single-particle $p$ shell. 

\emph{Attractive rank and Pauli blocking.---}
To characterize pair structures strong enough to matter, let $V_{\rm eff}$
denote the Hermitian two-neutron interaction in a chosen finite two-particle
space, and write its net attractive part as
\begin{equation}
W=[-V_{\rm eff}]_+
=\sum_\alpha g_\alpha|\phi_\alpha\rangle\langle\phi_\alpha|,
\qquad g_\alpha>0.
\label{eq:attractive_operator}
\end{equation}
For a chosen attraction threshold $\epsilon>0$, we define the attractive
rank $r_{\rm att}(\epsilon)$ as the number of eigenvalues of the interaction
kernel below $-\epsilon$, counting multiplicity. Weaker attractive modes are
not counted, and the strict matrix rank is recovered at $\epsilon=0$.
The partial-wave components contribute to the interaction kernel but are not
themselves the modes counted by $r_{\rm att}$; the threshold is applied only
after all attractive and repulsive contributions are combined.

We use the attractive rank as a conceptual tool for understanding how
sufficiently strong pair modes emerge in the low-lying valence space, rather
than as an additional quantity to be evaluated in the full large-scale NCSM
calculation. The restricted $0p_{3/2}$ analysis in the Supplemental Material,
Sec.~\ref{sec:p32_analysis}, provides a simple illustration.

For the rigorous rank-one limit, we take $\epsilon=0$ and suppose that only
one attractive pair state is present. Pauli blocking then gives
\begin{equation}
E_4^{(V)}-2E_2^{(V)}\ge0,
\label{eq:rank_one_bound}
\end{equation}
and for a reduced pairing Hamiltonian $E_4\ge2E_2$; proofs are given in the
Supplemental Material. Thus, repeated use of a single attractive pair
structure cannot provide more than additive interaction binding.

Additional attractive pair structures can evade this restriction. If an
added attractive interaction acts in a symmetry sector orthogonal to the
two-neutron ground state, it leaves $E_2$ unchanged up to the first level
crossing but can lower $E_4$. Hence $X=E_4-2E_2$ can decrease when the
four-particle state makes use of the additional sector. The relevant physical
question is therefore whether more than one sufficiently strong independent
pair structure is available. The same restricted $0p_{3/2}$ example underlying
Eq.~(\ref{eq:p32_restricted_balance}) makes this picture explicit. The $J=0$
and $J=2$ matrix elements, $V_0$ and $V_2$, separate the interaction gained
by adding the first and second valence-neutron pairs. Their partial-wave
decomposition shows that removing the $^3P_2$ contribution substantially
reduces the attraction in the net $J=2$ sector.

By contrast, reduced valence-space descriptions provide a useful comparison but do not
give a unique microscopic partial-wave assignment. The finite-range
Furutani--Horiuchi--Tamagaki (FHT) interaction
\cite{Furutani:1978FHT,Furutani:1979FHT}, used in the reduced form of
Ref.~\cite{Fossez:2018Helium}, contains several even-$L$ components after
projection, while fitted terms can absorb continuum coupling, core
polarization, and induced interactions. It can therefore reproduce helium
spectroscopy without uniquely identifying the microscopic partial waves
responsible for the extra quartet binding, whereas the \emph{ab initio}
Daejeon16 NCSM calculation permits that decomposition.

The observables play complementary roles; no single one is by itself a unique
signature of quartet formation. The negative $X_H$ establishes nonadditive
binding along the isotope chain. Within the neutron-neutron interaction, the
partial-wave decomposition identifies conventional $^1S_0$ pairing together
with additional $^3P_2$ attraction, while the restricted $0p_{3/2}$ example
shows how the $^3P_2$ contribution can make an additional net pair mode
sufficiently attractive. The irreducible three-subshell $^3P_2$ contribution
and the four-body density cumulants given in the Supplemental Material show
that the relevant four-neutron structure extends across multiple subshells
rather than being an inert $(0p_{3/2})^4$ shell closure. Taken together, these
observations provide complementary evidence for a cooperative mechanism of
quartet formation. The negative interaction contribution is largely offset by
a positive intrinsic-kinetic-energy contribution, leaving the three-point
energy difference near $-1$ MeV.

More generally, the thresholded attractive-rank picture provides a simple
organizing principle. Pauli blocking limits repeated use of one strong pair
mode, whereas additional independent modes passing the chosen threshold
provide other structures through which the four-neutron system can gain interaction
energy. We therefore interpret the four valence neutrons of $^8$He as an
interaction-driven bound two-pair structure consistent with the quartet
component of multimodal superfluidity, arising from the interplay of $^1S_0$
pairing and additional attraction in the $^3P_2$ partial wave.

This closes the picture with which we began. The second valence pair of
$^8$He is more bound than the first because the four neutrons do not simply
repeat one pair structure; Pauli blocking forbids that route from yielding
more than additive binding. The restricted $0p_{3/2}$ space already gives
the extra binding of the second pair, but not enough absolute binding to bind
the valence neutrons in either $^6$He or $^8$He. The irreducible multi-subshell
contributions and four-body cumulants show how the correlated four-neutron
structure extends into higher subshells and supplies the additional
correlations required in the full calculation. The attractive $^1S_0$
and $^3P_2$ neutron-neutron interactions work together to provide the extra
binding of the second pair of valence neutrons in $^8$He, a finite-system
counterpart of the quartet predicted for multimodal neutron superfluidity.
 
The underlying physics should apply
to other neutron-rich nuclei now becoming accessible at rare-isotope facilities. $^8$He is one of the simplest systems in which the answer can be
studied in detail from \emph{ab initio} calculations, and it suggests that
quartet correlations are not an accident, but the expected outcome
once more than one strong pair mode is available and other nuclear
structure effects are not masking the signal.

\emph{Data availability.---}
The data associated with this work are available at
\href{https://drive.google.com/drive/folders/1NxG_N7BbB_w6Dj_DXeGvpFOw6hXJuaZa?usp=sharing}{this Google Drive directory}.

\begin{acknowledgments}
We thank Ik~Jae~Shin for providing the $N_{\max}=12$ $^8$He results.  We gratefully acknowledge the help and guidance of Takayuki Miyagi and Calvin Johnson in using NuHamil and BIGSTICK codes. We also thank
Alex Brown, Witek Nazarewicz, Simin Wang, and members of the Nuclear Lattice Effective
Field Theory Collaboration for helpful discussions.  D.L. and Y.-Z.M.
acknowledge U.S. Department of Energy grants DE-SC0013365, DE-SC0023175,
DE-SC0026198, and DE-SC0023658; U.S. National Science Foundation grant
PHY-2310620; as well as Oak Ridge Leadership Computing Facility computing
resources through the INCITE award ``Ab initio nuclear structure and nuclear
reactions'' and Michigan State University's Institute for Cyber-Enabled
Research and High-Performance Computing Center. The work of Y.-H.S. was supported by
the Rare Isotope Science Project of the Institute for Basic Science
(IBS-I001-01) and the National Research Foundation of Korea (NRF), funded
by the Ministry of Science and ICT
(2013M7A1A1075764, RS-2024-00436392).
Computational resources were provided by the National Supercomputing Center
of Korea, including technical support (KSC-2024-CRE-0256).
\end{acknowledgments}

\bibliography{References}


\clearpage
\onecolumngrid

\begin{center}
{\large\bf Supplemental Material for\\[1mm]
``Evidence for Quartet Binding of Valence Neutrons in $^8$He''}
\end{center}

\vspace{4mm}

\setcounter{equation}{0}
\setcounter{figure}{0}
\setcounter{table}{0}
\setcounter{section}{0}

\renewcommand{\theequation}{S\arabic{equation}}
\renewcommand{\thefigure}{S\arabic{figure}}
\renewcommand{\thetable}{S\arabic{table}}
\renewcommand{\thesection}{S\arabic{section}}

\setcounter{secnumdepth}{1}

\def\theHequation{S\arabic{equation}}
\def\theHfigure{S\arabic{figure}}
\def\theHtable{S\arabic{table}}
\def\theHsection{S\arabic{section}}

\section{Basis-convergence results}

Tables~\ref{tab:binding_hw15} and \ref{tab:binding_hw25} give the
ground-state energies and three-point energy differences for the Daejeon16
calculations at $\hbar\omega=15$ MeV and $25$ MeV.

\begin{table}[h]
\centering
\caption{
Ground-state binding systematics for the helium isotopes using Daejeon16
at $\hbar\omega=15$ MeV. All energies are in MeV. Energy differences are
evaluated from the unrounded energies before the displayed values are
rounded. The $^8$He $N_{\max}=12$ result is from
Ref.~\cite{Shin:private}.
}
\label{tab:binding_hw15}
\begin{ruledtabular}
\begin{tabular}{ccccccc}
$N_{\max}$ &
$E_4$ &
$E_6$ &
$E_8$ &
$E_6-E_4$ &
$E_8-E_6$ &
$X_H$
\\
\hline
0  & -22.564 & -18.215 & -17.382 & +4.349 & +0.833 & -3.516 \\
2  & -25.702 & -23.707 & -22.983 & +1.996 & +0.724 & -1.272 \\
4  & -27.786 & -27.149 & -27.889 & +0.637 & -0.740 & -1.377 \\
6  & -28.218 & -28.367 & -29.583 & -0.149 & -1.216 & -1.067 \\
8  & -28.336 & -28.861 & -30.405 & -0.525 & -1.544 & -1.019 \\
10 & -28.365 & -29.095 & -30.787 & -0.730 & -1.692 & -0.962 \\
12 & -28.370 & -29.211 & -30.990 & -0.841 & -1.778 & -0.938 \\
\hline
Exp. & -28.296 & -29.271 & -31.396 & -0.975 & -2.125 & -1.150 \\
\end{tabular}
\end{ruledtabular}
\end{table}

\begin{table}[h]
\centering
\caption{
Ground-state binding systematics for the helium isotopes using Daejeon16
at $\hbar\omega=25$ MeV. All energies are in MeV. Energy differences are
evaluated from the unrounded energies before the displayed values are
rounded. The $^8$He $N_{\max}=12$ result is from
Ref.~\cite{Shin:private}.
}
\label{tab:binding_hw25}
\begin{ruledtabular}
\begin{tabular}{ccccccc}
$N_{\max}$ &
$E_4$ &
$E_6$ &
$E_8$ &
$E_6-E_4$ &
$E_8-E_6$ &
$X_H$
\\
\hline
6  & -28.292 & -26.842 & -26.909 & +1.450 & -0.067 & -1.517 \\
8  & -28.346 & -27.750 & -28.335 & +0.596 & -0.585 & -1.181 \\
10 & -28.362 & -28.279 & -29.218 & +0.083 & -0.940 & -1.023 \\
12 & -28.368 & -28.608 & -29.798 & -0.241 & -1.190 & -0.949 \\
\hline
Exp. & -28.296 & -29.271 & -31.396 & -0.975 & -2.125 & -1.150 \\
\end{tabular}
\end{ruledtabular}
\end{table}

At $\hbar\omega=25$ MeV the binding converges more slowly than at
15 MeV, but the same hierarchy appears. At $N_{\max}=10$,
$E_6-E_4=+0.083$ MeV while $E_8-E_4=-0.857$ MeV, giving
$X_H=-1.023$ MeV. At $N_{\max}=12$, the corresponding values are
$-0.241$ MeV, $-1.431$ MeV, and $-0.949$ MeV.

The dominant multi-subshell partial-wave result is also stable.
The aggregate three-subshell contributions to $X_{nn,V}$ change from
\begin{equation}
(-0.420,+0.010,-0.006,-0.018,-1.092)~\mathrm{MeV}
\end{equation}
for
$(^1S_0,^1D_2,^3P_0,^3P_1,^3P_2)$ at $N_{\max}=8$ to
\begin{equation}
(-0.295,+0.015,-0.004,-0.047,-1.034)~\mathrm{MeV}
\end{equation}
at $N_{\max}=10$. In both spaces the $^3P_2$ contribution is the
dominant irreducible three-subshell contribution.

\section{Intrinsic kinetic-energy and interaction balance}
To analyze the different contributions of Hamiltonian components, we measure the expectation values $E_{\mathcal{O}}$ with $\mathcal{O}\in\{T_{\mathrm{ref}}, V_{{nn}}, V_{{pn}}, V_{{pp}}, V, H\}$ according to Eq.~\ref{eq:expect} in the main text.
More specifically, for equal nucleon masses,
\begin{align}
T_{\rm rel}
&=
\sum_{i=1}^{A}\frac{\bm p_i^2}{2m}
-\frac{1}{2Am}
\left(
\sum_{i=1}^{A}\bm p_i
\right)^2
\\
&=
\frac{1}{2Am}
\sum_{i<j}
(\bm p_i-\bm p_j)^2.
\end{align}

The coefficient $1/A$ and the numbers of neutron-neutron,
proton-neutron, and proton-proton pairs all change along the helium
isotope chain. Pair-resolved three-point differences of $T_{\rm rel}$
therefore mix changes in relative-momentum matrix elements with kinematic
changes in pair counting. We consequently report $X_{T_{\rm rel}}$ only
for the complete intrinsic kinetic operator and resolve by pair type only
the potential energy.

\begin{table}[h]
\centering
\caption{
Energy balance for the $N_{\max}=10$ Daejeon16 calculation.
The intrinsic kinetic energy and full Hamiltonian are shown only as total
three-point differences, while the potential contribution is further
decomposed into neutron-neutron, proton-neutron, and proton-proton terms.
All entries are in MeV.
}
\label{tab:energy_balance}
\begin{ruledtabular}
\begin{tabular}{lc}
Quantity & Three-point difference \\
\hline
$T_{\rm rel}$ & $+10.817$ \\
$V_{nn}$      & $-5.795$ \\
$V_{pn}$      & $-5.187$ \\
$V_{pp}$      & $-0.797$ \\
$V$           & $-11.780$ \\
$H$           & $-0.962$ \\
\end{tabular}
\end{ruledtabular}
\end{table}

The small negative $X_H$ is the residual of two much larger full-system
contributions. Resolving the potential energy shows that the neutron-neutron
interaction contributes $X_{nn,V}=-5.795$ MeV, while the proton-neutron
interaction contributes $X_{pn,V}=-5.187$ MeV and compensates part of the
large positive intrinsic-kinetic-energy contribution. Because the core and
intrinsic relative motion reorganize along the isotope chain, we do not assign
the proton-neutron term a separate quartet mechanism.

\section{Attractive rank}

Consider a finite antisymmetric two-neutron Hilbert space $\mathcal H_2$ and
let $V_{\rm eff}$ denote the Hermitian two-neutron interaction represented in
that space. Its net attractive part is
\begin{equation}
W=[-V_{\rm eff}]_+
=\sum_\alpha g_\alpha|\phi_\alpha\rangle\langle\phi_\alpha|,
\qquad g_\alpha>0.
\label{eq:supp_W}
\end{equation}
For a chosen attraction threshold $\epsilon$, the attractive rank
$r_{\rm att}(\epsilon)$ counts the orthonormal eigenstates with
$g_\alpha>\epsilon$, including multiplicity; weaker attractive modes are not
counted. At $\epsilon=0$ this reduces to the ordinary matrix rank of $W$.
The partial-wave components contribute to $V_{\rm eff}$ but are not
themselves the eigenmodes counted by $r_{\rm att}$.

The threshold is applied only after all attractive and repulsive interaction
components are combined. Repulsion in the same two-particle subspace can
weaken or eliminate a nominally attractive mode, so ranks assigned to
separately labeled interaction pieces cannot in general be added. The strict
rank-one results below refer to the exact $\epsilon=0$ limit.

\section{Restricted $0p_{3/2}$ matrix-element analysis}
\label{sec:p32_analysis}

To connect the restricted energy balance in
Eq.~(\ref{eq:p32_restricted_balance}), the partial-wave decomposition in
Fig.~\ref{fig:partial_wave}, and the attractive-rank discussion, we consider
the same fixed-core $0\hbar\Omega$ case in which all valence neutrons are
restricted to the $0p_{3/2}$ subshell. This is smaller than the complete
$N_{\max}=0$ NCSM space, but the center-of-mass separation remains exact in
this special case. All retained configurations lie at the minimum total
oscillator quanta allowed by antisymmetry. Writing
$N_{\rm tot}=N_{\rm intr}+N_{\rm cm}$, an excitation with
$N_{\rm cm}>0$ would require $N_{\rm intr}$ below this minimum and is
therefore impossible. Thus every state in the restricted space has
$N_{\rm cm}=0$; restricting further to $0p_{3/2}$ removes states from this
sector but does not introduce center-of-mass admixtures.

Core-core terms are then constant and core-valence terms are affine in the
number of valence neutrons $n$, so their three-point differences vanish. The
kinetic contribution vanishes exactly as well. Since the center of mass
remains in its oscillator $0s$ state,
\begin{equation}
T_{\rm lab}(n)
=
\left(3+\frac{5n}{4}\right)\hbar\omega,
\qquad
T_{\rm cm}=\frac{3}{4}\hbar\omega,
\end{equation}
for $n=0,2,4$. Hence
\begin{equation}
T_{\rm rel}(n)
=
T_{\rm lab}(n)-T_{\rm cm}
=
\left(\frac{9}{4}+\frac{5n}{4}\right)\hbar\omega,
\qquad
X_{T_{\rm rel}}^{(0p_{3/2})}=0.
\end{equation}
The center-of-mass subtraction is therefore independent of $n$ and preserves
the affine dependence of the intrinsic kinetic energy. For two neutrons in
this subshell, the allowed pair angular momenta are $J=0$ and $J=2$. We define the corresponding diagonal
neutron-neutron matrix elements as $V_0$ and $V_2$. The
$(0p_{3/2})^2$, $J=0$ configuration has interaction energy $V_0$, while the
filled $(0p_{3/2})^4$, $J=0$ configuration has exact pair counts
$N_{J=0}=1$ and $N_{J=2}=5$, giving interaction energy $V_0+5V_2$.
Using isotope labels only as shorthand for these idealized restricted
configurations, rather than for the full NCSM wave functions, we have
\begin{equation}
V_{nn}(^6{\rm He})-V_{nn}(^4{\rm He})=V_0,\qquad
V_{nn}(^8{\rm He})-V_{nn}(^6{\rm He})=5V_2,
\end{equation}
and therefore
\begin{equation}
X_{nn,V}^{(0p_{3/2})}=5V_2-V_0.
\end{equation}
The negative value diagnoses the extra binding of the second valence pair
relative to the first. It does not imply that the restricted configurations
are themselves bound: with all valence neutrons confined to $0p_{3/2}$, the
idealized $^6$He and $^8$He configurations both remain above their respective
breakup thresholds.

\begin{figure}[h]
\centering
\includegraphics[width=0.58\textwidth]
{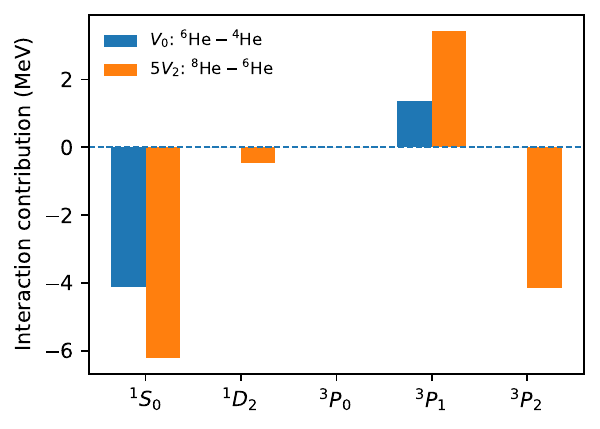}
\caption{
Partial-wave decomposition of the neutron-neutron interaction contribution
for adding the first and second valence-neutron pairs in the restricted
$0p_{3/2}$ space. The first-pair contribution is $V_0$, while the
second-pair contribution is $5V_2$.
}
\label{fig:p32_pair_addition}
\end{figure}

At $\hbar\omega=15$ MeV, the Daejeon16 neutron-neutron two-body matrix
elements give
\begin{equation}
V_0=-2.758~{\rm MeV},\qquad
V_2=-1.478~{\rm MeV},
\end{equation}
and hence
\begin{equation}
X_{nn,V}^{(0p_{3/2})}=-4.634~{\rm MeV}.
\end{equation}
These values are obtained directly from the restricted Daejeon16 two-body
matrix elements, not from a separate core-plus-four-neutron diagonalization.
They are also distinct from the isolated-$0p_{3/2}$ inclusion--exclusion
contribution in Fig.~\ref{fig:partial_wave}, which is evaluated with the
projected full NCSM wave functions. Table~\ref{tab:p32_partial} and
Fig.~\ref{fig:p32_pair_addition} show the partial-wave decomposition of the
restricted matrix elements. The $^1S_0$ interaction provides the dominant
attraction for the first pair and remains attractive for the second pair.
The $^3P_2$ interaction contributes only to $V_2$ in this restricted space
and provides a large additional attraction for the second pair, while
$^3P_1$ is repulsive.

\begin{table}[h]
\centering
\caption{
Partial-wave decomposition of the Daejeon16 neutron-neutron interaction
in the restricted $0p_{3/2}$ space. All entries are in MeV.
}
\label{tab:p32_partial}
\begin{ruledtabular}
\begin{tabular}{lrrr}
channel & $V_0$ & $5V_2$ & $5V_2-V_0$ \\
\hline
$^1S_0$ & $-4.125$ & $-6.197$ & $-2.073$ \\
$^1D_2$ & $ 0.000$ & $-0.477$ & $-0.477$ \\
$^3P_0$ & $ 0.000$ & $ 0.000$ & $ 0.000$ \\
$^3P_1$ & $+1.367$ & $+3.417$ & $+2.050$ \\
$^3P_2$ & $ 0.000$ & $-4.134$ & $-4.134$ \\
\hline
Total & $-2.758$ & $-7.392$ & $-4.634$ \\
Without $^3P_2$ & $-2.758$ & $-3.258$ & $-0.500$
\end{tabular}
\end{ruledtabular}
\end{table}

The same example illustrates the attractive-rank picture. Taking
$\epsilon=1$ MeV as a representative attraction threshold, the full
Daejeon16 interaction has both $V_0<-1$ MeV and $V_2<-1$ MeV. Counting
magnetic multiplicity therefore gives $r_{\rm att}=6$. If the $^3P_2$
interaction is removed while all other partial waves are kept unchanged,
$V_0$ remains $-2.758$ MeV but $V_2$ becomes
\begin{equation}
V_2=-0.652~{\rm MeV}.
\end{equation}
The $J=2$ eigenvalue is then above the $-1$ MeV cutoff and no longer
counts as sufficiently attractive, reducing $r_{\rm att}$ from $6$ to $1$. At the same time,
$X_{nn,V}^{(0p_{3/2})}$ changes from $-4.634$ MeV to only
$-0.500$ MeV. Thus the $^3P_2$ interaction both opens a strong additional
attractive pair sector and provides most of the nonadditive interaction
binding in this restricted example.

\section{Pauli blocking for a single pair state}

Suppose the net attractive two-body interaction contains only one
attractive pair state,
\begin{equation}
W
=
g|\phi\rangle\langle\phi|,
\qquad
g>0,
\qquad
\langle\phi|\phi\rangle=1.
\end{equation}
Define the pair creation operator by
\begin{equation}
P^\dagger|0\rangle
=
|\phi\rangle.
\end{equation}
The corresponding attractive interaction is
\begin{equation}
V_{\rm att}
=
-gP^\dagger P.
\end{equation}

Every normalized antisymmetric two-fermion state can be written in a
canonical pair basis. We therefore choose orthonormal single-particle states
$\mu,\bar\mu$ such that
\begin{equation}
P^\dagger
=
\sum_\mu
\sqrt{p_\mu}\,
a_\mu^\dagger a_{\bar\mu}^\dagger,
\qquad
p_\mu\ge0,
\qquad
\sum_\mu p_\mu=1.
\label{eq:canonical_pair}
\end{equation}
Define
\begin{equation}
D
=
\sum_\mu
p_\mu
\left(
a_\mu^\dagger a_\mu
+
a_{\bar\mu}^\dagger a_{\bar\mu}
\right)
\ge 0,
\end{equation}
where $D\ge0$ means that $D$ is positive semidefinite, or equivalently
has only nonnegative eigenvalues. The fermionic anticommutation relations
then give
\begin{equation}
[P,P^\dagger]
=
1-D,
\label{eq:pair_comm}
\end{equation}
and hence
\begin{equation}
PP^\dagger
=
P^\dagger P+1-D
\le
P^\dagger P+1.
\label{eq:operator_bound}
\end{equation}
For Hermitian operators $O$ and $O'$, the notation $O\le O'$ means that
$O'-O$ is positive semidefinite.

Let $\mathcal H_n$ denote the sector containing $n$ pairs, and let
$\lambda_n$ be the largest eigenvalue of $P^\dagger P$ on $\mathcal H_n$.
Since
\begin{equation}
P:\mathcal H_n\rightarrow\mathcal H_{n-1},
\end{equation}
suppose that $|\psi\rangle\in\mathcal H_n$ is an eigenstate of
$P^\dagger P$ with a nonzero eigenvalue $\lambda$,
\begin{equation}
P^\dagger P|\psi\rangle=\lambda|\psi\rangle,
\qquad \lambda>0.
\end{equation}
Then $P|\psi\rangle\neq0$, and
\begin{equation}
PP^\dagger(P|\psi\rangle)
=
P(P^\dagger P|\psi\rangle)
=
\lambda P|\psi\rangle.
\end{equation}
Thus every nonzero eigenvalue of $P^\dagger P$ in $\mathcal H_n$ is also
an eigenvalue of $PP^\dagger$ in $\mathcal H_{n-1}$. Conversely, if $|\chi\rangle\in\mathcal H_{n-1}$ satisfies
\begin{equation}
PP^\dagger|\chi\rangle=\lambda'|\chi\rangle,
\qquad \lambda'>0,
\end{equation}
then $P^\dagger|\chi\rangle\neq0$, and
\begin{equation}
P^\dagger P(P^\dagger|\chi\rangle)
=
P^\dagger(PP^\dagger|\chi\rangle)
=
\lambda' P^\dagger|\chi\rangle.
\end{equation}

Therefore the two operators
have the same nonzero eigenvalues in these two sectors.
 Therefore
\begin{equation}
\lambda_n
=
\lambda_{\max}\!\left(
PP^\dagger\big|_{\mathcal H_{n-1}}
\right).
\end{equation}
Using Eq.~(\ref{eq:operator_bound}), we have
\begin{equation}
PP^\dagger
=
P^\dagger P+1-D,
\end{equation}
with $D\ge0$. Therefore, for every state
$|\chi\rangle\in\mathcal H_{n-1}$,
\begin{equation}
\langle\chi|PP^\dagger|\chi\rangle
=
\langle\chi|(P^\dagger P+1)|\chi\rangle
-
\langle\chi|D|\chi\rangle
\le
\langle\chi|(P^\dagger P+1)|\chi\rangle.
\end{equation}
Taking the maximum over normalized states in $\mathcal H_{n-1}$ gives
\begin{equation}
\lambda_n
=
\lambda_{\max}\!\left(
PP^\dagger\big|_{\mathcal H_{n-1}}
\right)
\le
\lambda_{\max}\!\left[
\left(P^\dagger P+1\right)
\big|_{\mathcal H_{n-1}}
\right]
=
\lambda_{n-1}+1.
\end{equation}

For the one-pair sector, $P^\dagger P$ is the projector onto the normalized
pair state $|\phi\rangle=P^\dagger|0\rangle$, so
\begin{equation}
\lambda_1=1.
\end{equation}
Applying the preceding inequality for $n=2$ then gives
\begin{equation}
\lambda_2\le2.
\label{eq:lambda2}
\end{equation}

Let $E_N^{(V)}$ denote the lowest eigenvalue of $V_{\rm att}$ in
the $N$-particle sector. For one pair $E_2^{(V)}=-g$, whereas for two pairs
$E_4^{(V)}\ge-2g$. Thus
\begin{equation}
E_4^{(V)}
-
2E_2^{(V)}
\ge0.
\label{eq:rankone_pauli}
\end{equation}
A single attractive fermion-pair mode therefore cannot provide more than
additive interaction binding when occupied by two identical pairs. This is the familiar Pauli-blocking effect for composite fermion pairs
\cite{Richardson:1964Pairing,Rombouts:2002CompositeBoson,Pogosov:2010TwoPairs}.

Equation~(\ref{eq:rankone_pauli}) concerns the attractive interaction
operator itself. A corresponding statement for the complete energy follows
for a reduced pairing Hamiltonian. Let
\begin{equation}
b_i^\dagger
=
a_i^\dagger a_{\bar i}^\dagger,
\qquad
(b_i^\dagger)^2=0,
\end{equation}
and
\begin{equation}
H_{\rm red}
=
\sum_{ij}
h_{ij}
b_i^\dagger b_j.
\label{eq:reduced_pairing}
\end{equation}
For one pair, the Hamiltonian is the matrix $h$. If the $b_i$ were
unrestricted bosons, two pairs would both occupy the lowest eigenstate of
$h$, giving twice the one-pair ground-state energy. The physical
fermion-pair Hilbert space is the subspace satisfying the hard-core
constraint $(b_i^\dagger)^2=0$. Restricting the Hilbert space cannot lower
the variational ground-state energy, and therefore
\begin{equation}
E_4 \ge 2E_2
\label{eq:reduced_pair_convexity}
\end{equation}
for the reduced pairing Hamiltonian~(\ref{eq:reduced_pairing}).

\section{Monotonicity under an added attractive channel}

We now consider how the ground-state energy changes when the strength of an
attractive two-body interaction is increased. Let $B\ge0$ denote the
corresponding positive-semidefinite interaction, with its action understood
in the $N$-particle sector under consideration. We define
\begin{equation}
H_N(\lambda)
=
H_N(0)-\lambda B,
\qquad
\lambda\ge0,
\label{eq:Hlambda_supp}
\end{equation}
and let $E_N(\lambda)$ denote the ground-state energy of $H_N(\lambda)$.

If $\lambda_2\ge\lambda_1$, then
\begin{equation}
H_N(\lambda_2)
=
H_N(\lambda_1)
-
(\lambda_2-\lambda_1)B
\le
H_N(\lambda_1),
\end{equation}
because $B\ge0$. The variational principle therefore gives the exact
inequality
\begin{equation}
E_N(\lambda_2)
\le
E_N(\lambda_1).
\label{eq:variational_monotonicity}
\end{equation}
Thus increasing the strength of an attractive channel can only lower, or
leave unchanged, the ground-state energy.

At values of $\lambda$ where the ground-state energy is differentiable, the
Hellmann--Feynman theorem gives
\begin{equation}
\frac{dE_N}{d\lambda}
=
-
\langle
\Psi_N(\lambda)
|
B
|
\Psi_N(\lambda)
\rangle
\le0,
\label{eq:HF}
\end{equation}
where $|\Psi_N(\lambda)\rangle$ is the normalized ground state. The
inequality is strict whenever
\begin{equation}
\langle
\Psi_N(\lambda)
|
B
|
\Psi_N(\lambda)
\rangle
>
0.
\end{equation}
In other words, the energy decreases strictly whenever the ground state
actually uses the attractive channel.

The ground-state energy is also a concave function of $\lambda$. This
follows directly from the variational principle, since $E_N(\lambda)$ is the
minimum over normalized states of expectation values that are linear in
$\lambda$. When the ground state is nondegenerate, the same statement can be
seen from second-order perturbation theory:
\begin{equation}
\frac{d^2E_N}{d\lambda^2}
=
-2
\sum_{m\ne0}
\frac{
|\langle m|B|0\rangle|^2
}{
E_m-E_0
}
\le0.
\label{eq:concavity}
\end{equation}
Hence strengthening an attractive interaction produces a ground-state energy
that is both nonincreasing and concave as a function of its coupling
strength.

\section{Different symmetry sectors and quartet-selective binding}

We now ask when an attractive channel can lower the energy of four particles
without strengthening the two-particle ground state. Throughout this section, $B$
denotes the same two-body interaction; its action in the two- or
four-particle Hilbert space is inferred from context.

We first work in the two-particle Hilbert space. Let $W\ge0$ denote the attraction already present and let $B\ge0$
denote the added attractive contribution. If the two operators act in
different orthogonal symmetry sectors,
\begin{equation}
WB=BW=0,
\label{eq:orthogonal_channels}
\end{equation}
then the numbers of attractive states add:
\begin{equation}
\operatorname{rank}(W+\lambda B)
=
\operatorname{rank}W+\operatorname{rank}B,
\qquad
\lambda>0.
\label{eq:rank_add}
\end{equation}
Thus an attractive interaction in a new orthogonal two-particle sector
provides genuinely new attractive states.

Even without exact orthogonality, adding a positive-semidefinite attractive
operator cannot reduce the number of attractive states. Indeed,
\begin{equation}
\ker(W+B)=\ker W\cap\ker B.
\label{eq:kernel_sum}
\end{equation}
To see this, suppose $(W+B)|x\rangle=0$. Then
\begin{equation}
0
=
\langle x|W|x\rangle
+
\langle x|B|x\rangle.
\end{equation}
Both terms are nonnegative, so each must vanish separately. For a
positive-semidefinite operator this implies
\begin{equation}
W|x\rangle=B|x\rangle=0.
\end{equation}
Hence
\begin{equation}
\ker(W+B)=\ker W\cap\ker B\subseteq\ker W,
\end{equation}
and therefore, in a finite-dimensional two-particle space,
\begin{equation}
\operatorname{rank}(W+B)\ge\operatorname{rank}W.
\end{equation}
Thus the new attraction either leaves the number of attractive two-particle
states unchanged or increases it. For orthogonal attractive sectors, the
increase is exactly $\operatorname{rank}B$.

We next consider the case in which the added attraction acts in a
different symmetry sector from the two-particle ground state. Suppose the two-particle Hilbert space separates
into orthogonal symmetry sectors,
\begin{equation}
\mathcal H_2
=
\mathcal H_a\oplus\mathcal H_b\oplus\cdots,
\end{equation}
and the two-particle ground state belongs to $\mathcal H_a$. Let the new
attractive interaction act only in a different sector $\mathcal H_b$,
\begin{equation}
B=P_bBP_b,
\qquad
B\ge0,
\qquad
P_aB=BP_a=0,
\label{eq:sym_forbidden}
\end{equation}
where $P_a$ and $P_b$ project onto $\mathcal H_a$ and $\mathcal H_b$,
respectively.

For two particles,
\begin{equation}
H_2(\lambda)
=
H_2(0)-\lambda B.
\end{equation}
Because $B$ vanishes in the symmetry sector containing the two-particle
ground state, the energy of that state is unchanged. Therefore, until a
state from another symmetry sector crosses below it,
\begin{equation}
E_2(\lambda)=E_2(0).
\label{eq:E2fixed}
\end{equation}

The situation can be different for four particles. Although the same
two-body interaction $B$ is being used, a four-particle wave function can
contain pair components in the $\mathcal H_b$ sector even when the
two-particle ground state does not. Thus, for
\begin{equation}
H_4(\lambda)
=
H_4(0)-\lambda B,
\end{equation}
the monotonicity result above gives
\begin{equation}
E_4(\lambda_2)
\le
E_4(\lambda_1),
\qquad
\lambda_2\ge\lambda_1.
\end{equation}
Combining this with Eq.~(\ref{eq:E2fixed}), the quartet-binding measure
\begin{equation}
X(\lambda)
=
E_4(\lambda)-2E_2(\lambda)
\end{equation}
satisfies
\begin{equation}
X(\lambda_2)
\le
X(\lambda_1).
\label{eq:quartet_monotonicity}
\end{equation}
Thus attraction in a different two-particle symmetry sector, if it is
used by the four-particle state, can only make the quartet more bound relative
to two copies of the original two-particle ground state.

At a differentiable point, Eq.~(\ref{eq:HF}) gives
\begin{equation}
\frac{dX}{d\lambda}
=
-
\langle
\Psi_4(\lambda)
|
B
|
\Psi_4(\lambda)
\rangle,
\label{eq:dX}
\end{equation}
because $E_2(\lambda)$ is unchanged. Therefore
\begin{equation}
\frac{dX}{d\lambda}<0
\label{eq:strict_quartet}
\end{equation}
whenever
\begin{equation}
\langle
\Psi_4(\lambda)
|
B
|
\Psi_4(\lambda)
\rangle
>
0.
\end{equation}
In other words, if the four-particle ground state actually uses the new
attractive sector, increasing that attraction strictly enhances quartet
binding while leaving the two-particle ground state unchanged. This result is
nonperturbative up to the first symmetry-changing level crossing in the
two-particle system.

\section{Relation to previous pairing results}

Pauli blocking of repeated occupation of a single fermion-pair
state is well established. Richardson and Sherman obtained exact eigenstates of the
pairing-force Hamiltonian with the Pauli principle included explicitly
\cite{Richardson:1964Pairing}. Rombouts \emph{et al.} related the maximum
occupation of a composite boson state to the spectrum of a generalized
fermionic pairing problem \cite{Rombouts:2002CompositeBoson}. Pogosov,
Combescot, and Crouzeix solved the two-Cooper-pair problem for a separable
reduced BCS interaction and explicitly showed that Pauli blocking decreases
the binding per pair when the second pair is added
\cite{Pogosov:2010TwoPairs}.

The purpose of the present formulation is therefore not to identify Pauli
blocking as a new effect. Rather, the thresholded attractive rank provides a
compact way to count the sufficiently strong independent pair states of the
net two-neutron interaction. Combined with the exact monotonicity result for
attraction in a different symmetry sector, it provides an organizing
principle for the cooperative $^1S_0$ and $^3P_2$ structure found in the
Daejeon16 calculation.

\section{Selected four-body cumulants}

The energy observables establish nonadditive attraction between the valence
pairs. To verify independently that the wave function contains connected
four-neutron structure, we calculate diagonal four-body density cumulants
at $N_{\max}=10$. For single-particle states $a,b,c,d$,
\begin{align}
\rho^{(IV)}_{abcd}
={}&
\rho^{(4)}_{abcd}
-
\sum_{(2,2)}
\rho^{(II)}\rho^{(II)}
-
\sum_{(1,3)}
\rho\,\rho^{(III)}
\nonumber
\\
&-
\sum_{(1,1,2)}
\rho\rho\,\rho^{(II)}
-
\rho_a\rho_b\rho_c\rho_d.
\label{eq:cumulant}
\end{align}
Here $\rho^{(II)}$ and $\rho^{(III)}$ denote the connected two- and
three-body density cumulants, respectively. The sums run over the distinct
partitions of the four labels, subtracting all disconnected one-, two-, and
three-body contributions. The constrained
orbital cumulants are accumulated over magnetic substates as described in
Ref.~\cite{Ma:2026dti}.

Only the most diagnostic values are listed in
Table~\ref{tab:cumulants}. The isolated $0p_{3/2}$ four-body cumulant in
$^8$He is nearly zero, as expected for an independent filled subshell.
In contrast, the largest connected signal is the cross-subshell
$\{0p_{1/2},0p_{3/2}\}$ cluster, which is almost an order of magnitude
larger than the largest positive cumulant in $^6$He. Additional positive
signals connect the $p$ shell to the $sd$ shell, the radial $1p_{3/2}$
orbital, and the $pf$ shell.

\begin{table}[h]
\centering
\caption{
Selected positive four-neutron density cumulants in $^8$He at
$N_{\max}=10$, together with the isolated $0p_{3/2}$ value.
The last row gives the largest positive cumulant found in $^6$He for
scale comparison.
}
\label{tab:cumulants}
\begin{ruledtabular}
\begin{tabular}{lc}
Orbital cluster & $\rho^{(IV)}$ \\
\hline
$^8$He: $\{0p_{1/2},0p_{3/2}\}$
& $3.32\times10^{-2}$ \\
$^8$He: $\{0p_{3/2},0d_{5/2}\}$
& $4.41\times10^{-3}$ \\
$^8$He: $\{0p_{1/2},0p_{3/2},1p_{3/2}\}$
& $3.58\times10^{-3}$ \\
$^8$He: $\{0p_{1/2},0p_{3/2},0f_{5/2}\}$
& $2.36\times10^{-3}$ \\
$^8$He: $\{0p_{3/2}\}$
& $-8.70\times10^{-4}$ \\
$^6$He: $\{0p_{3/2},1p_{3/2}\}$
& $3.91\times10^{-3}$ \\
\end{tabular}
\end{ruledtabular}
\end{table}

The cumulant data are consistent with, but are not used as the primary
definition of, quartet binding. The strongest evidence is energetic:
$X_H<0$ is the residual of a large attractive interaction contribution
and a comparably large positive intrinsic kinetic contribution. Within the
neutron-neutron interaction, Daejeon16 shows cooperative $^1S_0$ pairing and
additional $^3P_2$ attraction, with $^3P_2$ dominant in the irreducible
multi-subshell contribution. This multi-subshell interaction signal and the
four-body cumulants together show that the relevant four-neutron structure
is not simply an inert $(0p_{3/2})^4$ determinant; the cumulants independently
demonstrate connected four-neutron correlations across multiple subshells.

\end{document}